\documentclass[reqno,tbtags,a4paper,11pt]{amsart}
\usepackage{epsfig}   
\usepackage{amsmath,amssymb,amscd,amsthm,a4wide}
\usepackage[mathscr]{eucal} 
\usepackage{calrsfs}  
\usepackage{graphicx} 
\usepackage[all]{xy}
\usepackage{a4wide}

\newtheorem{thm}{Theorem}

\newtheorem{lem}{Lemma}
\newtheorem{cor}{Corollary}

\theoremstyle{definition}
\newtheorem*{dfn}{Definition}

\newcommand{\Span}{\rm{Span}}
\newcommand{\CF}{\rm{CF}}
\newcommand{\res}{{\rm res}}

\newcommand\fA{{\mathfrak{A}}}

\usepackage{color}

\title
{Cyclic Frobenius algebras.}%
\author{V.~M.~Buchstaber$^\$$,  A.~V.~Mikhailov$^{\natural}$}
\address{$^\$$ Steklov Mathematical Institute, Moscow,\, Department of Mechanics and Mathematics, Moscow State University, Russia}
\email{buchstab@mi-ras.ru}
\address{$^\natural$ University of Leeds, Leeds, UK}
\email{a.v.mikhailov@leeds.ac.uk}
\thanks{AVM acknowledges support  from EPSRC Grant  EP/V050451/1.}

\begin{document}

\maketitle

Let $\mathcal{A}$ be an associative $\mathbb{C}$-algebra with identity 1, and let $\mathcal{M}$ be some $\mathbb{C}$-linear space, $\dim\mathcal{M}\geqslant 1$.
\begin{dfn}\label{D-1}
A  cyclic Frobenius algebra ($\CF$-algebra) $\mathcal{A}$ is an algebra $\mathcal{A}$
with a $\mathbb{C}$-bilinear skew-symmetric form
$\Phi(\cdot,\cdot) \colon \mathcal{A}\otimes_\mathbb{C} \mathcal{A}\to \mathcal{M}$ such that
$\Phi(A,BC)+\Phi(B,CA)+\Phi(C,AB)=0$ where $A,B,C \in\mathcal{A}$.
\end{dfn}

Examples: 1) $\mathcal{M}=\mathcal{A}$ and $\Phi(A,B)=AB-BA=[A,B]$;\, 2) $\mathcal{A}$ is a commutative algebra
with Poisson bracket $\{\cdot,\cdot\}$, $\mathcal{M}=\mathcal{A}$ and $\Phi(A,B)=\{A,B\}$;
3) $\CF$-algebras, where $\mathcal{M}$ is a field, obtained from the constructions of algebras from \cite{Ag-01}, \cite{ORS-12}.

Let us introduce the $\mathbb{C}$-linear subspace $\Span\subset\mathcal{A}$, spanned by all commutators $[A,B]\in \mathcal{A}$,
and the projection  $\pi\colon \mathcal{A}\to \mathcal{A}/\Span$.
Set $A\approx B$ if $A-B\in\Span$.

\begin{lem}\label{L-1}
Let $\varphi(\cdot,\cdot)\colon \mathcal{A}\otimes\mathcal{A}\to \mathcal{A}$ be a $\mathbb{C}$-bilinear skew-symmetric form
such that $\varphi(A,BC)\approx\varphi(A,B)C + B\varphi(A,C)$.
Then $\mathcal{A}$ is a $\CF$-algebra with $\mathcal{M} = \mathcal{A}/\Span$ and $\Phi = \pi\varphi$.
\end{lem}

Let $\mathcal{K} = \{ B\in\mathcal{A}\,:\,\Phi(A,B) = 0, \text{ for all}\, A\in\mathcal{A} \}$.
From Definition \ref{D-1} it follows that $\mathcal{K}$ is the subring in $\mathcal{A}$, $1\in \mathcal{K}$, and
$\Phi(A,bC) = \Phi(Ab,C)$ for all $A,C \in\mathcal{A}$ and $b\in\mathcal{K}$.

Let $\fA = \mathbb{C}\langle u_0,u_1,\ldots\rangle = \oplus\fA_k,\, k\geqslant 0$, be  a free associative graded algebra
with derivation $D$, $|u_k| = k+2$, $D(u_k)=u_{k+1},\, k=0,1,\ldots$\,.
We introduce the algebra $\fA^D = \{ A = \sum_{i\leqslant m}a_iD^i,\, a_m\neq 0,\, m\in \mathbb{Z},\, a_i\in \fA_{|a_m|+m-i} \}$,
where $[D,u_k] = u_{k+1},\; [D^{-1},u_k] = \sum_{i\geqslant 1}(-1)^i \\u_{k+i}D^{-i-1}$.
Let $A_+ + A_- = A \in \fA^D$, where $A_+ = \sum a_iD^{i},\, 0\leqslant i\leqslant m$,
for $m\geqslant 0$, and $A_+ = 0$, for $m<0$. We have $\res[D,A] = D(\res A)$, where $\res A = a_{-1}$.

We introduce a homogeneous $\mathbb{C}$-bilinear skew-symmetric form $\sigma(\cdot,\cdot)\colon \fA^D\otimes\fA^D \to
\fA$, $|\sigma(A,B)| = |A|+|B|$, by
\[
\sigma(aD^n,bD^m)\!=\!
\begin{cases}
\frac{1}{2}\binom{n}{n+m+1}\!\sum\limits_{s=0}^{n+m}\!(-1)^s \big(a^{(s)}b^{(n+m-s)}\!+b^{(n+m-s)}a^{(s)}\big),
\text{for } n+m \geqslant 0,\, nm<0,\\[1pt]
0,\qquad \mbox{otherwise}.
\end{cases}
\]

\begin{lem}\label{L-2}
For any $A,B \in \fA^D$, we have $D\big(\sigma(A,B)\big) = \res\,[A,B] - \text{$\Delta$}(A,B)$, where
\[
\Delta(aD^n,bD^m) =
\begin{cases}
\frac{1}{2}\binom{n}{n+m+1}\Big([a,b^{(n+m+1)}] + (-1)^{n+m}[b,a^{(n+m+1)}]\Big), \text{ for } n+m \geqslant -1,\\[1pt]
0,\qquad \mbox{otherwise}.
\end{cases}
\]
\end{lem}

\begin{lem}\label{L-3}
Under the canonical projection $\pi\colon \fA\to \fA/\Span = \mathcal{M} = \oplus\mathcal{M}_k,\; k\geqslant 0$, the operator $D$
gives monomorphisms $\overline{D}\colon \mathcal{M}_k \to \mathcal{M}_{k+1},\; k>0$.
\end{lem}

\begin{thm}\label{T-1}
The algebra $\fA^D$ is a $\CF$-algebra with the form $\Phi = \pi\sigma\colon \fA^D\otimes\fA^D\to \mathcal{M}$ such that $\mathcal{K} = \fA$ and $\Phi(D^n,D^{-n}) = n,\;n\in \mathbb{Z}$.
\end{thm}

\begin{cor}
On the set of generators $\{D^n,\, n\in \mathbb{Z}\}$ of the free left $\fA$-module $\fA^D$, the form $\Phi$ is non-degenerate.
\end{cor}

Let $L = D^2-u$ and $\mathcal{L} = D + \sum I_{i}D^{-i},\; i\geqslant 1,\, I_{i}\in \fA_{i+1}$, where
$\mathcal{L}^2 = L$. We obtain a sequence of series $\mathcal{L}^{2k-1},\, k\in\mathbb{N}$.
Set
\begin{equation}\label{eq-2}
\sigma_{2k-1,2n-1} = \sigma(\mathcal{L}_+^{2k-1},\mathcal{L}^{2n-1})\in\fA_{2n+2k-2}, \qquad
\rho_{2n} = \sigma_{1,2n-1} = \res\,\mathcal{L}^{2n-1},\, n>0.
\end{equation}
From the properties of the form $\sigma(\cdot,\cdot)$, we obtain $\sigma_{2k-1,2n-1} = \sigma_{2n-1,2k-1},\; k,n\in\mathbb{N}$.

Let us introduce derivations $D_{2k-1},\, k\in\mathbb{N}$, of the algebra $\fA$ such that
\begin{equation}\label{eq-3}
D_1 = D;\quad [D,D_{2k-1}] = 0;\quad D_{2k-1}(u) = -2D(\rho_{2k}),\; k\in\mathbb{N}.
\end{equation}

\begin{cor}
$[D_{2k-1},D_{2n-1}] = 0$.
\end{cor}
Let $u = u(t_1,t_3,\ldots)$. Set $\partial_{t_{2k-1}}(u) = D_{2k-1}(u)$.
\begin{thm}\label{T-2}
The system of equations $\partial_{t_{2k-1}}(u) = -2D\big(\rho_{2k}(u)\big),\, k\in \mathbb{N}$, where $\rho_2(u) = -\frac{1}{2}u$,
coincides with the KdV hierarchy on $\fA$
\[
4\partial_{t_3}(u) = D(u_2-3u^2);\quad 16\partial_{t_5}(u) = D(u_4-5u_2u-5uu_2-5u_1^2+10u^3);\, \ldots\,.
\]
\end{thm}
The {\it proof} follows from the above constructions (cf. \cite{B-Mikh-21}--\cite{Sok-20}).

Following \cite{B-Mikh-21-arXiv}, for $N\in \mathbb{N}$, set
$F_{2N+2}(u) = \rho_{2N+2} + \sum_{k=0}^{N-1}\alpha_{2(N-k+1)}\rho_{2k}$,
where $\rho_0=1$ and $\alpha_4,\ldots,\alpha_{2N+2}$ are free parameters.
Equation $F_{2N+2}(u) = 0$ is called the $N$-th  Novikov equation. Let $J(F)$
be the two-sided $D$-differential ideal in $\fA$ generated by the polynomial
$F_{2N+2}(u)$. Since $2^{2k+1}\rho_{2k+2} = u_{2k} + f(u_{2k-2},\ldots,u)$ and
$u_{k+1}=D^k(u)$, on the factor-algebra $\fA\diagup J(F) = \mathbb{C}\langle u, \ldots, u_{2N-1}\rangle$,
the KdV hierarchy (see Theorem 2) reduces to the $N$-th Novikov hierarchy, where the first system represents the $N$-th Novikov equation in the form
$D(u_k)=u_{k+1}$ for $0\leqslant k< 2N-1,\ D(u_{2N-1})=-f(u_{2k-2},\ldots,u)$.
It was shown in \cite{B-Mikh-21-arXiv} that, in terms of the form $\Phi$, (see Theorem 1), the polynomials
\begin{equation}\label{eq-4}
H_{2n+1,2N+1} = \sigma_{2n+1,2N+1} +
\sum_{k=1}^{N-1}\alpha_{2(N-k+1)}\sigma_{2n+1,2k-1},\; n=1,\ldots,N,
\end{equation}
define first integrals $\widehat{H}_{2n+1,2N+1}=\pi (H_{2n+1,2N+1})$ of the $N$-th Novikov hierarchy:
\[\pi (\partial_{t_{2k-1}}(\widehat{H}_{2n+1,2N+1}))=0,\qquad k,n\in\{1,\ldots,N\}.\]

\begin{thm}\label{t3}
In the quantum case, the $N$-th Novikov hierarchy (see \cite{B-Mikh-21-arXiv}) is written in the form of compatible systems of Heisenberg equations.
The polynomials \eqref{eq-4} are quantum commuting Hamiltonians of the hierarchy and are self-adjoint if the variables $u_k$ are self-adjoint and the values of the parameters $\alpha_4,\ldots,\alpha_{2N+2}$ are real.
\end{thm}

We would like to thank S.P. Novikov and V.N. Rubtsov for stimulative discussions of the results of this paper.

\end{document}